# Thermal stability study of nitrogen functionalities in a graphene network

Ajay Kumar[1,2], Abhijit Ganguly[2] and Pagona Papakonstantinou

Engineering Research Institute (ERI), School of Engineering, University of Ulster, Newtownabbey BT37 0QB, UK

E-mail: ajayepph@gmail.com and p.papakonstantinou@ulster.ac.uk

**Abstract**
Catalyst-free vertically aligned graphene nanoflakes possessing a large amount of high density edge planes were functionalized using nitrogen species in a low energy $N^+$ ion bombardment process to achieve pyridinic, cyanide and nitrogen substitution in hexagonal graphitic coordinated units. The evolution of the electronic structure of the functionalized graphene nanoflakes over the temperature range 20–800 °C was investigated *in situ*, using high resolution x-ray photoemission spectroscopy. We demonstrate that low energy irradiation is a useful tool for achieving nitrogen doping levels up to 9.6 at.%. Pyridinic configurations are found to be predominant at room temperature, while at 800 °C graphitic nitrogen configurations become the dominant ones. The findings have helped to provide an understanding of the thermal stability of nitrogen functionalities in graphene, and offer prospects for controllable tuning of nitrogen doping in device applications.

S Online supplementary data available from stacks.iop.org/JPhysCM/24/235503/mmedia

(Some figures may appear in colour only in the online journal)

A promising approach for tuning and controlling the electronic properties of graphene is doping of the hexagonal carbon network with heteroatoms, in a similar way to that elaborated in the silicon-based technology. Specifically doping with nitrogen atoms allows graphene transformation into an n-type semiconductor accompanied by the opening of a band gap. Controlled nitrogenation of graphene has been reported to improve device performance for biosensing, energy conversion and storage, field emission and nanoelectronics applications [1–8]. For example nitrogen-doped graphene, which is a metal-free catalyst, exhibits excellent electrocatalytic activity for oxygen reduction reaction (ORR) due to the induction of a net positive charge on carbon atoms as a result of the electron transfer between the adjacent carbon atoms and the N atoms in the conjugated graphene $\pi$ system [9]. Additionally, these nitrogen electrochemically active centres induce electron-donor properties on the carbon surface that give rise to pseudocapacitance through Faradaic interactions between the ions of electrolytes and the carbon electrode surface [10].

Nitrogen in graphene exhibits a very broad range of structures such as graphitic-like, pyridine-like, pyrrolic, cyanide etc structures. Different nitrogen sites play different roles in tailoring the structure of graphene and hence its performance. Recent experimental and theoretical work clarified that substitutional nitrogen doping increases the field emission (FE) performance by providing sufficient channels around the Fermi level, whereas the pyridine-like N doping reduces the FE [11, 12]. N-doped graphene is much more efficient for anchoring molecules as compared to pure graphene. A theoretical study using first-principles density functional calculations revealed that the pyridine-like nitrogen significantly strengthens metal adsorption compared to the substitutional nitrogen [13]. Understanding and controlling the chemistry and electronic structure in nitrogen-doped graphene is of foremost importance in achieving its demonstration in broad practical applications.

Nitrogen doping of graphene is an active area of research that is rapidly growing [14, 15]. Several approaches have been successfully demonstrated, even though the most common ones are through thermal annealing of graphene oxide in ammonia [16] or *in situ* doping during the chemical vapour deposition of graphene employing an $NH_3$ gas precursor [17]. Low energy nitrogen ion irradiation is a useful tool for doping the carbon network with controlled doses [18–21].

---
[1] Present address: Department of Chemical Engineering, Case Western Reserve University, Cleveland, OH 44106-7217, USA.
[2] These authors have contributed equally to this work.

In this work, the changes induced in the electronic structure of graphene nanoflakes (GNFs) by low energy nitrogen ion bombardment and the evolution of the nitrogen groups during vacuum thermal treatment have been studied using *in situ* high resolution x-ray photoemission spectroscopy. Since different bonding configurations have different bonding energies, it is thus possible to drive the system into a more prominent single-bonding configuration by thermal activation. Our studies established that at high temperatures, nitrogen in substitutional graphitic sites is the predominant moiety.

GNF samples were grown on a metal-catalyst-free Si wafer using a Seki Technotron microwave plasma enhanced chemical vapour deposition (MPECVD) system, using a $CH_4:N_2$ (1:1) reactive gas mixture (see the supplementary material for details available at stacks.iop.org/JPhysCM/24/235503/mmedia). High resolution x-ray photoelectron spectroscopy (XPS, ESCA300 spectrometer, Daresbury) was performed for determining the binding configurations and the chemical state of the elements present in the *in situ* $N^+$ ion bombarded GNFs and thermally annealed ones. The bombardment was performed at room temperature with 2 keV N ions at a dose of $10^{15}$ ions $cm^{-2}$ employing a low energy ion source VG AG21 for 10 min in the preparation chamber of the ESCA300 spectrometer. Heat treatment of the samples at temperatures of 400 °C, 600 °C and 800 °C was performed in the preparation chamber employing an electron beam heater with electron beam impact on the back surface of the Si substrate. The samples were cooled before collecting the spectra in the ultrahigh vacuum chamber.

Surface morphology of the as grown GNFs is shown in figure 1(a), while the high resolution TEM image (figure 1(b)) confirms the multilayered structure with $d_{spacing}$ of 0.3368 nm. *In situ* wide energy survey (WES) and high resolution elemental scans (C 1s, N 1s and O 1s) were taken at room temperature after each heat treatment. No additional signal except those of C, N and O was observed in the wide energy survey scan for all the samples. Figure 2(a) shows the atomic concentrations (at.%) of the $N^+$ ion bombarded GNF sample as estimated from the C 1s, N 1s and O 1s core levels under different conditions (table S1 available at stacks.iop.org/JPhysCM/24/235503/mmedia). A large drop in the nitrogen concentration was observed from 9.59 to 5.56 at.% with incrementation in the temperature from room temperature (RT) to 800 °C, while the carbon concentration showed a small enhancement from 89.26 to 94.23 at.%. Noticeably, a small amount of oxygen was also observed, which was decreased from 1.16 at.% at RT to an almost negligible amount of 0.21 at.% at 800 °C. The substantial drop of nitrogen content (>40%) with temperature implied self-arrangement or temperature-favoured competition between the different carbon–nitrogen bonding structures.

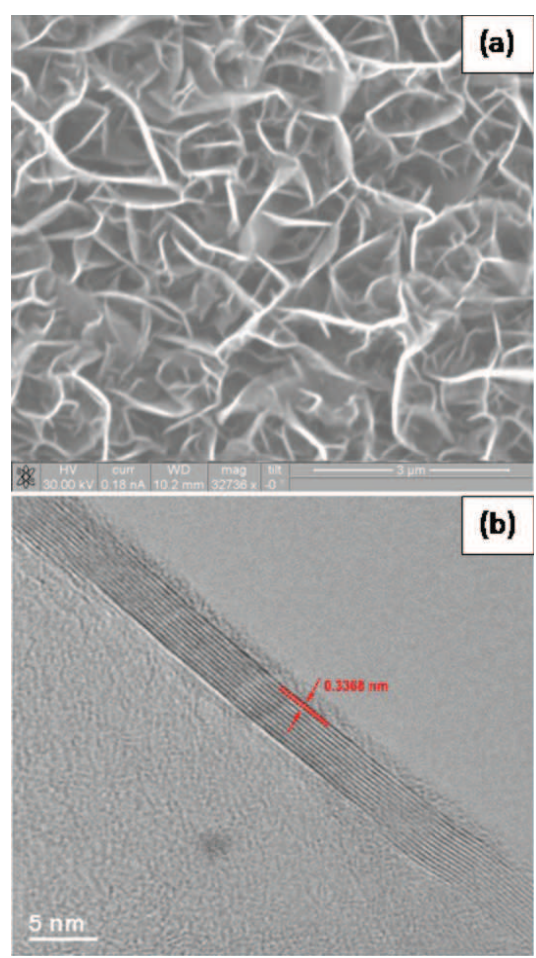

**Figure 1.** (a) SEM image showing the vertical orientation and sharp edges and (b) HRTEM images to show the multilayered structure of the pristine GNFs.

Figure 2(b) shows C 1s core level electron photoemission spectra for the *in situ* $N^+$ bombarded GNF sample as a function of annealing temperature. The deconvolution of the C 1s core level spectra was performed by fitting a form, after Shirley background subtraction (using the CASAXPS software provided with XPS system), with the lowest chi-square ($\chi^2$) value. The deconvoluted C 1s core level form (for RT; figure 2(b), and figure S1 available at stacks.iop.org/JPhysCM/24/235503/mmedia) consists of the five peaks C1 (284.5 ± 0.05 eV), C2 (285.2 ± 0.15 eV), C3 (286.6 ± 0.4 eV), C4 (288.01 ± 0.8 eV) and C5 (290.2 ± 0.2 eV). Upon judicious review of the available literature, we have assigned the C1 and C2 peaks as the $sp^2$-C aromatic carbon and defect peaks respectively [22–25]. The defect peak corresponds to $sp^3$ C–C bonds and is representative of fluctuations in the graphene structure, such as edges, branching and bending. It is also associated with the presence of amorphous carbon (a-C) on the graphene surface as revealed in the high resolution TEM cross sectional image of figure 1. The C3 peak is assigned to the nitrogen incorporation and C–O bonds [22, 26, 27], and the C4 peak to C=O bonds [28] as represented in figure 3. This assignment is in agreement with theoretical calculations that predict a chemical shift of ∼1–1.5 eV to higher BE relative to the $sp^2$ peak for the C–O bond [29]. The C=O double-bond emission occurs at an even higher BE range. The last peak, C5, component mainly arises from $\pi \rightarrow \pi^*$ transitions, a characteristic of aromatic C structure, and partly from the C=O or adsorbed oxygen components. It is believed that the contribution of O moieties to the C5 component at RT is more pronounced compared to that at the higher temperatures; however, due to practical limitations in the deconvolution process the estimation of each contribution is restricted (figure S2 available at stacks.iop.org/JPhysCM/24/235503/mmedia). Consequently, subsequent thermal annealing, at 400 °C, resulted in a slight decrease in C5 intensity, inevitably due to the removal of adsorbed oxygen moieties, as observed in figure 5(a) (and figure S3 available at stacks.iop.org/JPhysCM/24/235503/mmedia). However, with the increase in annealing temperature from 400 to 800 °C, a slight but steady increase of the C5 component could be observed, signifying the restoration of the delocalized $\pi^*$ conjugation. The development of a more symmetric C1 peak at 800 C is indicative of the $sp^2$ content increasing with the annealing. This incrementation in the $sp^2$ structure is at the cost of the depletion in the N atoms and oxygen contaminants with annealing.

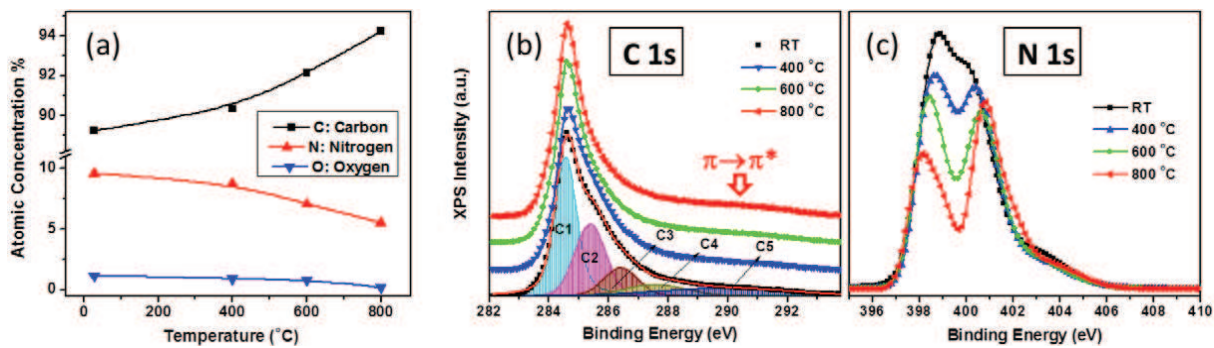

**Figure 2.** (a) C, N and O at.% concentration of the *in situ* $N^+$ ion bombarded GNF sample with heat treatment at 400, 600 and 800 °C. (b) C 1s and (c) N 1s core level photoemission spectra with the annealing temperature. Deconvoluted components of the C 1s (b) core level at room temperature are shown: experimental raw data (black symbols), the fitted results (red line) and subpeaks (coloured shades).

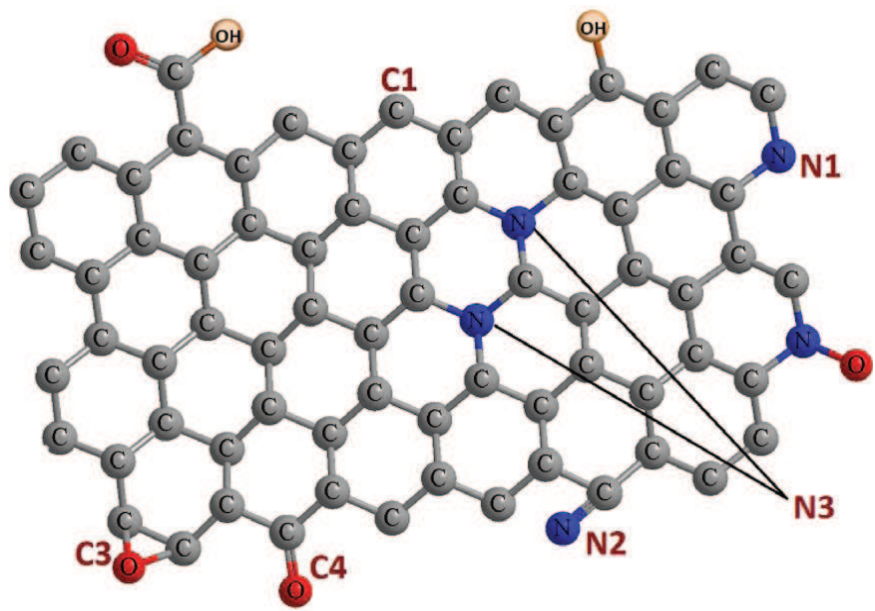

**Figure 3.** Schematic structure of nitrogen-doped graphene produced by low energy $N^+$ bombardment.

The deconvolutions of the C 1s core level photoemission spectra are tabulated in table S2 (available at stacks.iop.org/JPhysCM/24/235503/mmedia). The chemical shift in the core level spectra here can be attributed to the electronegativity of the functional group/atom and the C–C/C–N bond length, which also results in electronic perturbation of the core electrons [30]. Above 400 °C the $sp^2$ C1 peak increased with increasing annealing temperature due to the progressive removal of N and O moieties. In contrast, the C2 defect peak decreased in the same temperature range (400–800 °C) due to the concurrent removal of amorphous phase.

The effect of the annealing temperatures is more evident from the decrease in N concentration as illustrated in figure 2(c). The pristine *in situ* $N^+$ ion bombarded GNF sample shows a single broad asymmetric N 1s spectrum (denoted as RT) with dominant lower energy component, while the thermal annealing results in a double-peak formation indicating the ascendancy of lower energy components. Deconvolution of N 1s spectra revealed five kinds of carbon–nitrogen bonding structures at RT (figure 4), here assigned as N1, N2, N3, N4 and N5. The peak N1 is assigned to pyridine–N: substitutional tetrahedral N bonded to two C atoms in six-membered rings at the edge of the graphene layer; in this structure carbon and nitrogen are bonded with a double bond [22, 31, 32]. Peak N2 is attributed to the cyanide functional group (C ≡ N: the N atom gets bonded to a C atom without replacing it). Peak N3 represents graphite-like substitutional structure, in which N is bonded to three C atoms in the aromatic graphene structure [22, 31, 30]. Peaks N4 and N5 are assigned to the pyridine-*N*-oxide and chemisorbed nitrogen oxide species respectively [33, 34]. The peaks were fitted to Voigt functions having 80% Gaussian and 20% Lorentzian character, after performing a Shirley background subtraction as shown in figure 4. The deconvoluted parameters are illustrated in figure 5(b) (and also in figure S4 available at stacks.iop.org/JPhysCM/24/235503/mmedia). It was found that at the higher annealing temperatures (600 and 800 °C), N5 peak fitting was not possible due to surface oxygen removal from the network.

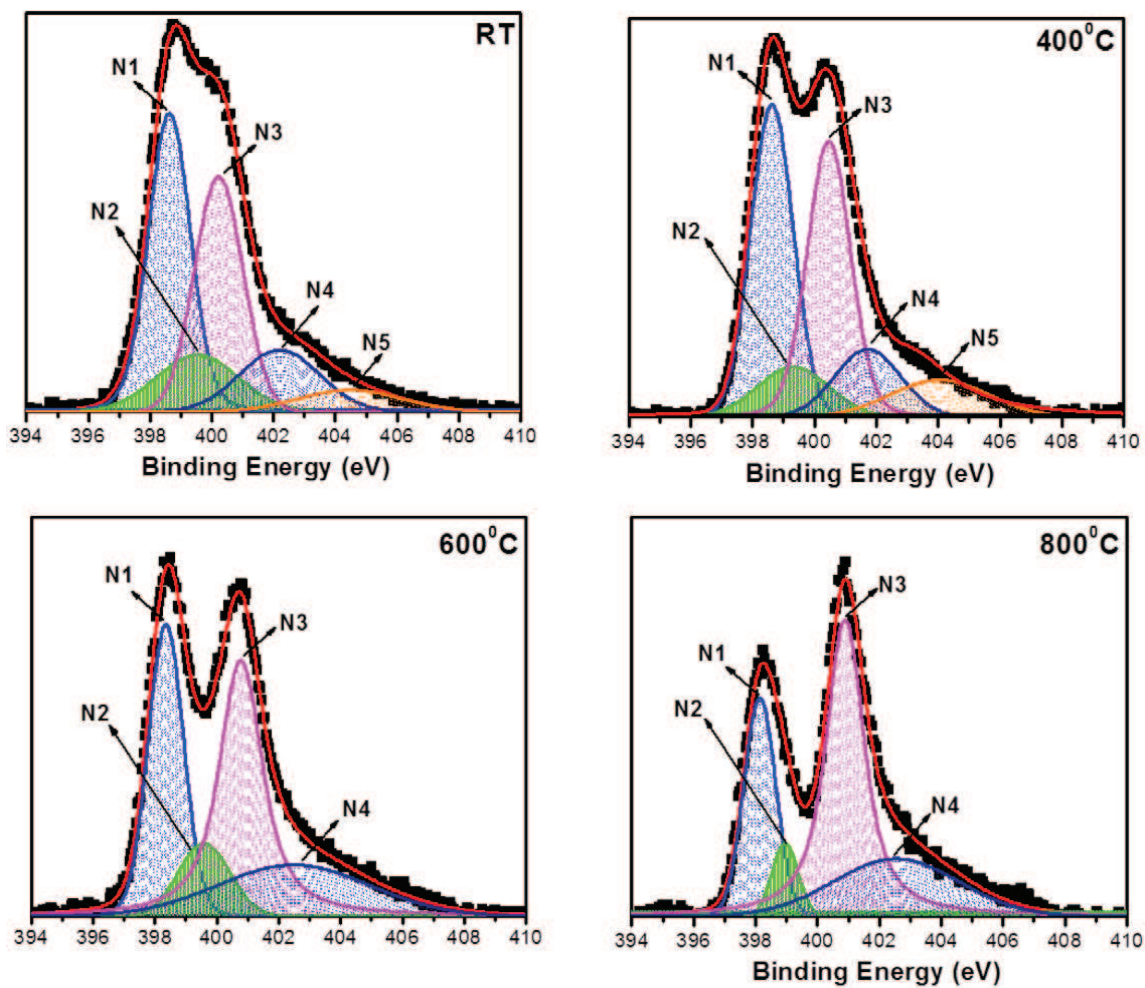

**Figure 4.** Deconvoluted components of N 1s core level photoemission spectra for (a) the *in situ* room temperature $N^+$ ion bombarded GNF sample, room temperature RT; and after heat treatments at (b) 400 °C, (c) 600 °C and (d) 800 °C.

Figure 5(b) shows the reduction of the absolute intensities of all the deconvoluted components of N 1s core level spectra with the annealing temperature. It was found that the graphitic N component (N3 peak) remains stable above 400 °C. Further analysis on the relative intensity of N components (relative to the total area under the N 1s core level spectra, ignoring the N5 peak due to the convoluted effects of N and O moieties on it) revealed an interesting effect of the annealing temperature as shown in figure 5(c). The pristine $N^+$ ion bombarded GNF sample showed the highest contribution of the pyridinic component (N1 peak), while the thermal annealing above 400 °C caused the dominance of the N3 graphitic N component. It was found that with the annealing temperature increase, there is a significant increase in the N3 peak (above 400 °C): this observation can explain the stable thermal nature of the graphitic N component [35, 36]. A decrease in the N1 peak implies a reduction in the pyridine-like nitrogen sites, while increase in the N3 peak intensity confirms the temperature-favoured evolution of substitutional graphitic N at the expense of pyridinic N1 components. Figure 5(d), illustrating the variation of the peak intensity of N components relative to the N3 peak, supports the above-mentioned supposition. In a similar manner, figure S4 (available at stacks.iop.org/JPhysCM/24/235503/mmedia) illustrates the similar trends of variation of peak intensities of

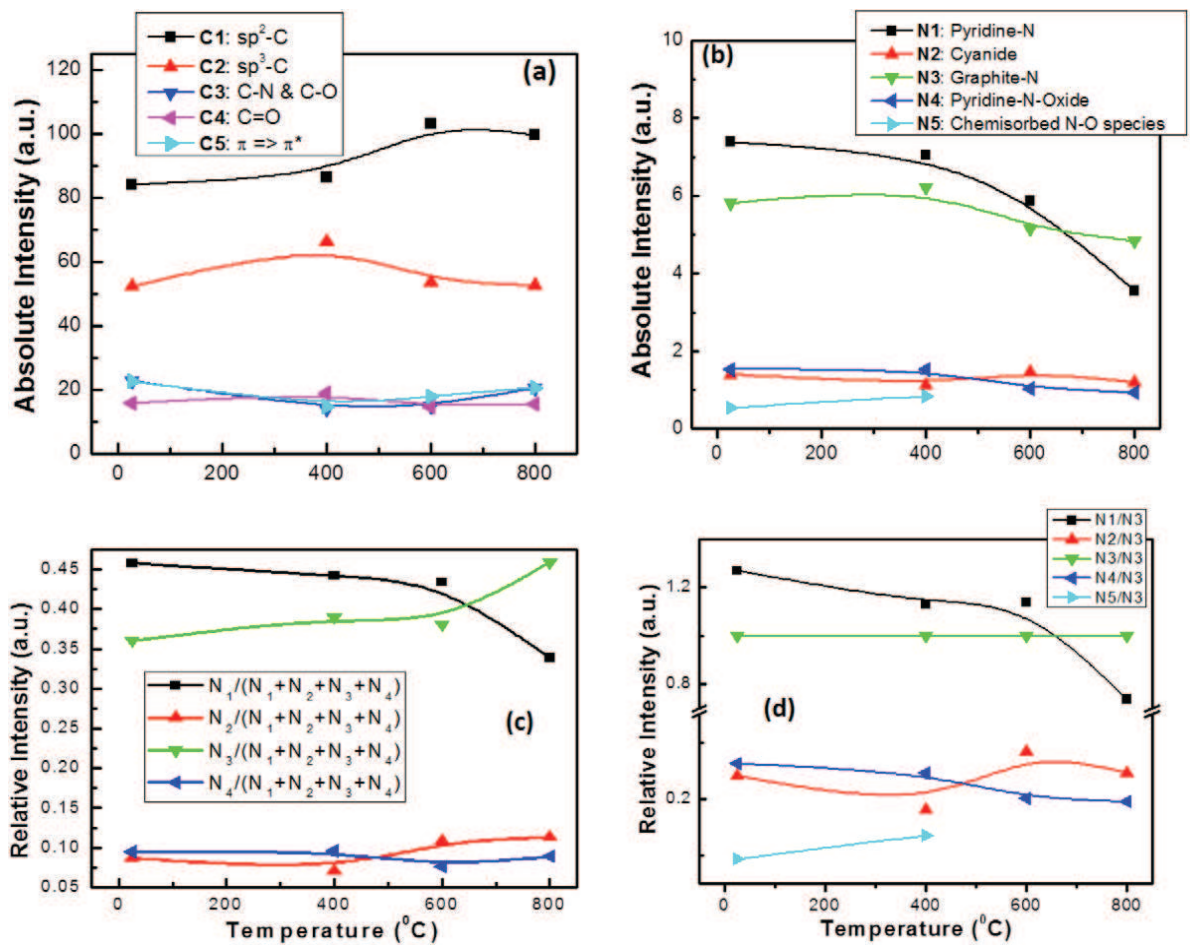

**Figure 5.** Effect of the annealing temperature on the (a) absolute peak intensities of C 1s deconvoluted components, and (b) absolute peak intensities of N 1s deconvoluted components, and those relative to the (c) total contribution of the N 1s core level spectra and (d) graphitic–N component assigned to peak N3.

N 1s components with the annealing temperature relative to other N 1s components.

The N5 peak is attributed to the convoluted effects of N and O moieties: chemisorbed N–O species. This phenomenon originates from the removal of N oxidic species by thermal treatment, which can be confirmed from the decreasing oxygen atomic concentration as shown in figure 2(a) and table S1 (available at stacks.iop.org/JPhysCM/24/235503/mmedia), and also from the decrease in the O 1s core level photoemission spectra with annealing temperature (figure S5 available at stacks.iop.org/JPhysCM/24/235503/mmedia).

In conclusion, we have studied the evolution of the electronic structure of the functionalized graphene nanoflakes over the temperature range 20–800 °C using *in situ* high resolution x-ray photoemission spectroscopy. We demonstrate that low energy irradiation is a useful tool for achieving nitrogen doping levels up to 9.6 at.%. Pyridinic configurations are found to be predominant at room temperature, while at 800 °C graphitic nitrogen becomes dominant. The findings have helped us to understand the thermal stability of nitrogen functionalities in graphene and they offer prospects for controllable tuning of nitrogen doping in device applications.

## Acknowledgments

This work was supported by a Vice-Chancellor Research Scholarship from the University of Ulster (Ajay Kumar), The Leverhulme Trust (Fellowship to Dr Ganguly: 1-212-R-0197), and EPSRC funded facility access to the NCESS XPS facility (EP/E025722/1) at Daresbury.